\documentclass{aastex}
\tighten 

\begin{document}

\received{}

\title{Tomographic Separation of Composite Spectra. \\
XI The Physical Properties of 
the Massive Close Binary HD~100213 (TU Muscae)} 

\author{Laura R. Penny\altaffilmark{1} and Cynthia Ouzts\altaffilmark{2}} 
\affil{Department of Physics and Astronomy, College of Charleston, 101 Science Center, \\
 58 Coming Street, Charleston, SC 29424; pennyl@cofc.edu}
\author{Douglas R. Gies\altaffilmark{1}}
\affil{Center for High Resolution Astronomy and \\
 Department of Physics and Astronomy, \\
 Georgia State University, P.O. Box 4106, Atlanta, GA 30302-4106; \\
 gies@chara.gsu.edu}

\altaffiltext{1}
{Guest Observer with the {\it International Ultraviolet Explorer} satellite.} 
\altaffiltext{2}
{Current address: Christ Church Episcopal School, 245 Cavalier Drive, Greenville, SC 29607; ouztsc@cces.org}

\slugcomment{to be submitted to ApJ}


\begin{abstract}

We present the results of a Doppler tomographic reconstruction of 
the UV spectra of the double-lined, O-binary \objectname[TU Muscae]{HD 100213} based on observations made 
with the {\it International Ultraviolet Explorer} ({\it IUE}).  
We used cross-correlation methods to obtain radial velocities, 
confirm the orbital elements, estimate the UV flux ratio, and 
determine projected rotational velocities. 
The individual component spectra are classified as O7~V $+$ O8~V 
using UV criteria defined by Penny, Gies, \& Bagnuolo.  
We present a model fit of the eclipsing light curve from observations from the {\it HIPPARCOS} satellite and published observations of Andersen \& Gronbech. We derive an orbital inclination, $i =77\fdg7 \pm 1\fdg0$.  
This analysis indicates that both stars are currently
experiencing Roche lobe overflow (RLOF), which confirms earlier results that this is one of
only a few massive contact binaries.  Our derived masses, $M_p/M_\odot = 16.8 \pm 0.4$ and $M_s/M_\odot = 10.5 \pm 0.3$, are significantly lower than those computed from the Doppler shifts of lines in the optical spectrum.  We suggest that the difference occurs because mutual irradiation decreases the upper atmospheric temperature gradient in the inwards facing hemispheres of both stars, which makes lower excitation lines appear weaker ther and shifts their center-of-light away from the center-of-mass.  We compare the current state of HD~100213 with predicted outcomes of massive close binary evolutionary models, and suggest that the system is currently in a very slow Case AA mass transfer stage.
\end{abstract}



\section{Introduction}

In this series of papers we have utilized the tomography algorithm to better study the individual stars that make up multiple systems.  Our targets 
have included: AO Cas \citep{Bag91}, $\phi$ Per \citep{Tha95}, HD~152248 \citep{Pen99}, HD~135240 \citep{Pen01}, and $\delta$ Ori \citep{Har02}.  This time we turn our attention to the double-lined, eclipsing, massive binary system, HD~100213 \citep[TU Mus; O8.5 V,][]{Wal71}.  This binary was originally studied by \citet[AG75]{And75} who presented a combined spectroscopic and photometric solution and concluded that the stars were in contact. Later \citet[ST95]{Sti95} performed a spectroscopic study using 24 high-resolution UV spectra from the {\it International Ultraviolet Explorer} ({\it IUE}).  Although their orbital period agreed well with that from AG75, their orbital semiamplitudes ($K$) were significantly smaller, resulting in smaller binary separation and masses.  \citet[T03]{Ter03}, with an eye towards resolving this discrepancy, obtained new optical ($\lambda\lambda3900-9200~\AA$) spectra at two orbital phases close to quadrature, as well as new photometric observations.  These they combined with the earlier spectroscopic and photometric data of AG75 and {\it HIPPARCOS} observations.  The results of this combined analysis supported the AG75 solution and appeared to rule out the smaller amplitude values from ST95. As both ST95 and T03 note, that this discrepancy between UV and optically determined $K$ values is not unique to TU MUS, but also is seen in several other massive semi-detached and contact binaries.

\citet[L07]{Lin07} obtained optical spectra of TU Mus to determine whether the system displayed the Struve-Sahade (SS) effect, i.e., a change in the relative line depths of the components between quadrature phases.  While they did not observe this effect, they did note several issues that are illuminating.  First, their orbital elements (specifically semiamplitudes) vary depending upon which lines were used, i.e., the $K$ values from \ion{He}{2} lines were smaller than those from \ion{He}{1}.  Second, the equivalent widths of two \ion{He}{1} lines are phase locked with the orbital motion.  They are at maximum strength when the back (opposite the companion) of the star is facing towards us and at a minimum half a orbit later.  They conclude that the \ion{He}{1} lines preferentially form over a portion of the stellar surface, and possibly an analogous, but opposite, effect may be occuring with the UV lines, due to mutual irradiation of the inner sides of the stars.

As one of only a few contact massive binaries, the TU Mus system is an important test case for interactive binary evolution models.  It is critical that well-determined values for the current separation and component masses be determined. In this paper we present our own analysis of the {\it IUE} observations of HD~100213 resulting in a double-lined orbital solution (\S 3), our tomographic reconstruction of the composite spectra into separate primary and secondary spectra and their respective spectral classifications
(\S 5), individual projected rotational velocities (\S 4), and our model of the eclipsing light curve (\S 6).  We discuss the discrepancy between the orbital semiamplitudes determined from optical spectra versus those from UV spectra in \S 7.  We compare the individual masses based on the spectroscopic and photometric orbits to those predicted by theoretical methods and discuss their implications (\S 8).


\section{Observations and Reductions}

There are 24 high dispersion, short wavelength prime camera 
spectra of HD~100213 available from {\it IUE}.  
These spectra were obtained from Multi-Mission Archive at Space Telescope Science Institute ({\it MAST}). One spectrum, SWP 54386 was found to be flawed, with several large gaps in spectral coverage. ST95 note that this spectrum was severely overexposed due to a satellite tracking problem.  This spectrum was not used for our analysis. The individual SWP image numbers and heliocentric Julian dates of mid-exposure of the remaining 23 observations are presented in Table~1.   

\placetable{t1}

The spectra were manipulated in several stages 
to produce a matrix of spectra (in dimensions of 
wavelength and time) rectified using a common set 
of relatively line-free zones and sampled with a uniform 
$\log \lambda$ wavelength grid.   The major interstellar 
absorption lines were replaced by straight line segments 
in the processing.  Details are given in \citet*{Pen97}.


\section{Radial Velocities and Orbital Elements} 

Radial velocities and orbital elements based on this same set 
of spectra were presented by ST95. However, we decided to 
check their results using our suite of cross-correlation 
techniques to obtain radial velocities and estimates of 
projected rotational velocities and flux ratio \citep[for details see][]{Pen97}.  We used two spectra (SWP22108, SWP37429) of HD~34078 \citep[O9.5V; 
projected rotational velocity $V\sin i = 26.5$ km~s$^{-1}$][]{Pen96a} 
for cross-correlation with the spectra of HD~100213.  The relative velocities were determined by fitting two Gaussians to the composite cross-correlation functions (ccfs).  
The widths of the two Gaussians and their intensity ratio were 
determined by least-squares fits of ccfs with well-separated peaks, i.e., those obtained within $\pm 0.12$ of the quadrature phases.  The final radial velocities (Table~1) were transformed from relative to absolute by adding the radial velocity of HD~34078 \citep[+54.4 km~s$^{-1}$, ][]{Gie87}.  \citet{Qia07} investigated possible cyclical variations in orbital period of TU Mus and suggested that there is a tertiary companion.  The predicted low mass of this star, $\approx 2.0 M{_\odot}$, indicates that it will not significantly contribute to the {\it IUE} spectra.  In addition there is a well known faint companion, with a magnitude $V=13.4$, that is located 15$\arcsec$ to the southwest of TU Mus, but we see no evidence of any tertiary spectrum in our ccfs.

We determined the orbital elements
using the program (SBCM) of \citet{Mor74}.  This program is limited to individual solutions for the component stars. We assigned zero weight to those velocities obtained from spectra within about 0.15 phase of the minima of eclipse and 1.0 weight on all other observations.  The orbital elements based on the {\it IUE} measurements are presented in Table~2.  The combined primary and secondary solutions of ST95 and T03 are presented along with the separate primary and secondary solutions from this paper.  We note that T03 performed a simultaneous solution with both the spectroscopic and photometric data using the 2003 Wilson-Devinney (WD) code.  They do not list their resulting $K$ values, but do present their final mass ratio, period, inclination, and separation, from which we determined their individual semiamplitudes (shown in Table 2).  The numbers in parentheses refer to the errors in the last digit quoted.   We ran the program holding the eccentricity at zero, and allowed the other parameters to vary.  We reached excellent agreement with ST95 in all parameters.    The radial velocity curves based on 
the data are illustrated in Figure~1.  The velocity residuals and orbital phases from the individual fits are listed in Table~1.  The epoch $T_0$ corresponding to zero phase corresponds to the radial velocity maximum of the primary star.

\placetable{t2}

\placefigure{f1}

Our results are not in agreement with those from T03 who find significantly higher orbital semi-amplitude velocities for both stars. As noted above the disagreement between orbital semi-amplitude values determined from UV spectra and those from optical spectra is not unique to TU Mus.  Four other massive semi-detached or contact systems have been spectroscopically studied in both the UV and optical: LY Aur \citep[HD~35921,][]{Sti94,May68,And74,Pop82}, $\delta$ Ori A \citep[HD36486,][]{Har02}, V Puppis \citep[HD~65818,][]{Sti98,And83}, and LZ Cep \citep[HD~209481,][]{How91,Har98}. All the Roche filling components have lower derived $K$ values from the ultraviolet analysis when compared to the optical. We suggest that the difference between optical and UV semi-amplitude values for TU Mus originates in the close proximity of the two stars.  In order to determine quantitatively how the tidal distortions affect the measured radial velocities, we need to complete the light curve analysis. From this analysis we will have a complete picture of the amount of distortion present. We save further discussion on the disagreement of optical and UV semi-amplitude values for \S 7 below.

\section{Projected Rotational Velocities} 

We use a method developed previously \citep{Pen96a} to estimate the individual projected rotational velocities of the component stars from their cross-
correlation functions (ccfs) with a narrow lined star.  For HD~100213, we use HD~34078 as our template star.  In the study above, we calibrated the relationship between the ccf Gaussian width, and $V \sin i$ using the \citet{Con77a} data sample.  The resulting calibration curve is given by
\begin{equation}
V \sin i = -3.830 \times 10^{-3} \sigma ^2 + 2.7903 \sigma - 80.7, \end{equation}
The primary's and secondary's Gaussian widths of $\sigma = 164.4 \pm 5.0$ and $131.3 \pm 5.0$ km~s$^{-1}$ correspond to a projected rotational velocities of $V\sin i = 274.5 \pm 8.0$ km~s$^{-1}$ and $219.6 \pm 8.0$ km~s$^{-1}$.  These agree within errors to those values determined by ST95 ($250 \pm 25$ and $195 \pm 20$ km~s$^{-1}$) and AG75 ($285 \pm 29$ and $240 \pm 24$  km~s$^{-1}$).  T03 do not quote errors on their $V\sin i$ values of $291$ and $242$ km~s$^{-1}$, respectively.

\section{Tomographic Reconstruction and Spectral Classification}

The separate primary and secondary star spectra can be derived using a 
tomography algorithm to reconstruct the individual spectra from 
the ensemble of composite spectra.  The algorithm is described in detail in a separate paper \citep[see][]{Bag94}. 
We have commented on some limitations of the algorithm previously \citep{Tha95}, 
but we note here that strong wind features, such as the P Cygni lines, 
may be reconstructed incorrectly since their radial velocity curves 
can be very different from those associated with the stars themselves.  
Since the tomographic reconstruction is based on orbital Doppler shifts, 
the reconstruction will be ambiguous in the vicinity of such wind features.

The tomographic reconstruction of the primary and secondary spectra is based
on the assumed flux ratio which may not be well determined at the outset.  The composite spectra are first separated using the ccf intensity ratios of $r_{ccf} = 0.49$ as an initial estimate of the UV flux ratios.  The resultant individual stellar spectra are then classified.  For the spectra of the primary and secondary we use the spectral diagnostics listed below.  We can then determine the true UV intensity ratio by numerical tests involving the cross-correlation of the template with a simulated binary spectrum formed from spectra of stars of the same spectral types as the primary and the secondary, with a range of input flux ratios \citep[for details see][]{Pen97}.   For the case of TU Mus (O7 V + O8 V), our measured ccf intensity ratio is $0.49 \pm 0.05$ using the template star, HD~34078 (O9.5 V). We found that we could match this ratio by combining the spectra of the single stars HD~36879 (as the primary, O7 V) and HD 48279 (as the secondary, O8 V) using a UV flux ratio, $r_{UV} = 0.48 \pm 0.05$. 

Our method for estimating the spectral types and luminosity classes 
of the separated primary and secondary spectra is based on the equivalent width measurements of several UV absorption lines \citep[for details see][]{Pen96b}.  We identify the two stars as O7 V (O6.5 -- O7.5) and O8 V (O7.5 -- O9) based on an evaluation of the following criteria:  for the primary the equivalent widths of four lines (\ion{Si}{3} $\lambda 1299$,  \ion{He}{2} $\lambda 1640$, \ion{Fe}{4} $\lambda 1681$, and \ion{Fe}{4} $\lambda 1765$); and for the secondary the same set plus \ion{Fe}{5} $\lambda 1429$ and the line ratio of \ion{He}{2} $\lambda 1640$/\ion{Fe}{5} $\lambda 1429$.  These classifications are, for the most part, consistent with Walborn's (1971) composite classification of O8.5 V.  \citet{Hil87} estimated the types as O7.8 + O8.2, which again agree well with our determinations. These spectral types correspond to temperatures \citep{Mar02} of $T_{\rm eff,p} = 37.2 \pm 1.5$ kK and $T_{\rm eff,s} = 34.7 \pm 1.5$ kK.


\section{Light Curve Analysis and Masses}

The {\it HIPPARCOS} light curve \citep{Esa97} shows two equally spaced minima of what appears to be $0.^{m}54 -- 0.^{m}48$ in {\it HIPPARCOS} magnitude.  We obtained, in electronic form, the AG75 differential photometric data from J. V. Clausen and J. Andersen.  These observations show that in fact the primary eclipse is slightly deeper, $0.58$ magnitude (in $V$).  Both sets of data also have ellipsoidal variations present between eclipses due to the tidal distortion of the stars inside the Roche surfaces.  We note that the time of minimum light for the {\it HIPPARCOS} observations is shifted in phase by 0.028 compared to that of AG75.  This is part of a larger trend examined by \citet{May04} and \citet{Qia07} suggesting that the period of TU Mus is lengthening due to slow mass transfer from the secondary to the primary. For the purpose of simultaneously fitting the {\it HIPPARCOS} and AG75 data, we calculated the phases of the photometric data using our final spectroscopic period and the time of primary minimum from the individual data sets themselves (i.e., the {\it HIPPARCOS} epoch of minimum light was taken as the time of the faintest observation).  The differential $V$ band photometry of AG75 was transformed to {\it HIPPARCOS} magnitude system (almost identical to $V$ for hot stars) and are plotted in Figure 2. Both the shapes and depths of the eclipses and the ellipsoidal variations are dependent upon the stellar radii and orbital inclination.  

We used the light curve synthesis code GENSYN \citep{Moc72} to produce $V$ differential light curves. As T03 mention there is a faint visual companion to the system 15$\arcsec$ to the SW which was included in both the {\it HIPPARCOS} and AG75 photometry.  The visual magnitude of the companion from the Guide Star Catalog is $V=13.46$. This represents $0.8\%$ of the measured light, which we removed before fitting the light curve.  The orbital parameters were taken from our above solution, and the physical fluxes and limb darkening coefficients were taken from \citet{Kur94} and \citet{Wad85}, respectively.  We neglect any treatment of radiation pressure following the suggestion of \citet{How97} .  Rotation rates were determined using the model inclination and radii plus the projected rotational velocities given above. 
Making a fit of the observed light curve requires estimates of many 
parameters, and initially we decided to constrain all the orbital parameters according to our spectroscopic orbit, set the 
stellar temperatures according to the spectral classifications ($T_{\rm eff,p} = 37.0$~kK, $T_{\rm eff,s} = 35.0$~kK), and determine the ratio of radii 
from the visual flux ratio, $F_s/F_p$ (determined from the method described in Penny et al. 1997 to be $0.49 \pm 0.05$).  Then, each model light curve is a function of the remaining two parameters, inclination and primary polar radius.  
There are three characteristics of the light curve that we sought to match: eclipse depth, eclipse duration, and amplitude of the ellipsoidal variation.  In order to fit the observed difference in depth between the primary and secondary eclipse, the primary's (secondary's) temperature was increased (decreased) to its highest (lowest) possible value, $T_{\rm eff,p} = 38.7$~kK, $T_{\rm eff,s} = 33.2$~kK. The critical issue in matching eclipse depths is the temperature difference between the two stars, not their actual values. (Of course the actual values of $T_{\rm eff}$ do have a large effect upon the calculated absolute magnitude of the system.)  The ellipsoidal variations are a sensitive function of assumed stellar radii, we found that models with the correct ellipsoidal variation depth {\it and} eclipse duration and depth had inclinations $76\fdg7 <i< 78\fdg7$. The best-fit model light curve (for $i=77^\circ$.7) is illustrated in Figure~2. In this model the stars are just in contact, both filling their Roche surfaces.  The larger inclination value of $78\fdg7$ fit the eclipse depths very well, but the ellipsoidal variations were too small.   Below $76\fdg7$ the widths of the eclipses become too broad and shallow.  At the higher inclination, both stars are just inside their Roche volumes and at the lower, they are in slight overcontact.  The presence of the ellipsoidal variations in conjunction with eclipses make this inclination determination very precise. 

\placefigure{f2}

Our best-fit inclination value agrees extremely well with the inclination presented by T03, who find an $i = 77\fdg8 \pm 0\fdg1$.  We caution against their small quoted errors.  No doubt these are statistical in nature, however they can be misleading.  To demonstrate we plot the secondary eclipse with our nominal best fit model ($i = 77\fdg7$), and our maximum and minimum inclination model fits ($i_{max} = 78\fdg7$, $i_{min} = 76\fdg7$) in Figure 3. All three fit the data very well.  Owing to the smaller $K$ values from our spectroscopic orbit, our estimates for the resulting separation and radii are quite different from that of T03.   The radii at the best fit inclination are $R_p/R_\odot = 7.2 \pm 0.5$ and $R_s/R_\odot = 5.7 \pm 0.5$.  Based on our $V\sin i$ values, the equatorial rotation speeds of the two stars are $281 \pm 15$ km s$^{-1}$ and $225 \pm 15$ km s$^{-1}$. The synchronous rotation rates, calculated using the equatorial radii, are $261 \pm 8$ and $208 \pm 7$ km~s$^{-1}$, for the primary and secondary, respectively.  These agree within our errors, and both stars are rotating at or slightly above synchronous.  To determine the distance to TU Mus we use the combined absolute magnitude of the pair from our light curve model, $M_V = -4.6 \pm 0.1$.  We note that T03's use of absolute magnitudes from a calibration of spectral class to absolute magnitude is far less accurate.  In fact their own light curve analysis produces $\log L/L_{\odot}$ values ($4.8 \pm 0.2$ and $4.5 \pm 0.2$) that are incompatible with their adopted absolute magnitudes (which correspond to $\log L/L_{\odot}$ values of $5.3$ and $4.9$).  We adopt a visual absorption value of $1.^m2$ and an apparent visual magntude of $8.^m23$ in accordance with AG75.  We determine a distance of $2.1 \pm 0.1$ kpc, which agrees within errors with that from AG75, but which is much smaller than $4.8$ kpc found by T03.  Our distance places the system in the nearby portion of the Carina spiral arm \citep{Rus03}.  

\placefigure{f3}

\section{Tidal and Irradiation Effects on Spectral Lines} 

From our combined spectroscopic and photometric analysis we have a complete model of the TU Mus binary.  We can now attempt to quantify why the optical and UV radial velocities result in disparate $K$ values.  We considered several effects.  The first is the shift of the the stars' centers of light from their centers of mass.  We used GENSYN to produce synthetic line profiles of typical UV and optical lines at photometric phase = 0.25, with a center of mass velocity of zero.  At this orbital phase, the absolute values of the radial velocities of the component stars should be equal to their semi-amplitude values.  To create model line profiles, GENSYN requires unbroadened line profiles for the both the primary and secondary.  We adopt \ion{He}{1} $\lambda 4471$ as our typical optical line.  The synthetic input flux profiles were obtained from the OSTAR2002 grid of \citet{Lanz03} using effective temperatures from our best fit light curve model, $T_{\rm eff,p} = 38700$~K, $T_{\rm eff,s} = 33200$~K, and $\log g = 4.0$ for both stars.  Our measured UV radial velocities are not from individual line profiles, but from a cross correlation of the spectrum of HD~34078 with the observed binary spectra.  We cross-correlated UV spectra from OSTAR2002 with our template HD~34078 and used the resulting ccfs as our input UV line profiles. Radial velocities were obtained by fitting double Gaussians to the inverted UV and optical GENSYN created profiles and are presented in Table 3.  Neither the UV nor the optical appear to be significantly affected by the non-sphericity of their shape (Case 1). The resultant velocities although slightly smaller are well within the measuring error of the expected semi-amplitude values.  

\placetable{t3}

Next we considered how the inner hemispheres of the components are heated by irradiation from the companion star, by setting the number of reflection iterations in GENSYN to equal five (Case 2).  This had again a small effect on the measured radial velocities and not in a manner that addresses the $K$ disparity, as both the UV and the optical lines are shifted towards the binary center of mass.  This is not surprising given that at both wavelengths higher temperatures mean increased luminosity and a greater shift of the line center. Next we considered whether absorption lines formed primarily in hotter regions, such as UV lines, would more preferentially form in these inner hotter hemispheres, resulting in lower $K$ values.  An additional input to GENSYN does allow us to specify how the primary and secondary line profiles vary with increasing temperature.  We again used the OSTAR2002 grid to quantify how the strength of the input lines change with temperature and we created an additional set of binary synthetic profiles with all three effects (Roche geometry, irradiation, line variability). A temperature increase of 1000 degrees is expected to cause a decrease in \ion{He}{1} $\lambda 4471$ equivalent width of 9\% while the predicted change is a decrease of 3\% for the UV ccfs. Disapointingly, the changes are too small to alter significantly the measured radial velocities.

One last effect we consider is how the heating of the inner hemisphere reduces the temperature gradient in the upper atmosphere.  Line depth is a function of the difference in temperature between the depths where the continuum and the line are formed, and in general, the larger the difference, the deeper the line. UV lines are high excitation features that are formed at depths comparable to the continuum.  Optical lines, on the other hand, are lower excitation features that form at much higher levels.  But in cases where irradiation from above occurs, the temperature where optical lines form will be higher than normal, thus reducing the line contrast. In this situation, optical lines formed on the inner, illuminated hemisphere will appear weaker while lines formed on the outer hemisphere have normal strength, and this shifts the effective line center away from the center of mass, leading to larger $K$ values. This effect would not so greatly change the UV line contrast as they are already formed at temperatures close to that of the continuum depth.  We suggest that this may be the cause of the larger semi-amplitude values from optical lines.  Without an accuate model of how irradiation changes the temperature gradient in the upper atmosphere it is difficult to quantify this effect.  However in a previous study on the Struve-Sahade (SS) effect, \citet{Gie97b} investigated the effects of increased wind-blanketing on the temperature gradient, and subsequent line strengths, for the secondary star in AO Cas.  Lines from transistions that tend to decrease in strength with temperature were not as strong, while those transitions that increased with temperature were either unchanged or intensified.  While the UV and \ion{He}{2} lines are somewhat affected by the increase in external heating, the largest impact is in the \ion{He}{1} features.  The largest effect can be seen in their Figure 5, where the \ion{He}{1} $\lambda 6678$ feature is drastically diminished.   In their case, they argue that a colliding bow shock region near the surface of secondary, creates heating on the approaching hemisphere of that star which is seen as the SS-effect.  For our situation, only the inner regions of the stars would be heated from irradiation.

We find observational support for this hypothesis in the recent paper by L07 in their measured equivalent width values for two \ion{He}{1} lines, $\lambda 4026$ and $\lambda 4471$.  In both cases the lines variations are phase locked, with maximum values at phases when we see the outer hemisphere of the stars.  This ia consistent with the idea that the optical lines formed where irradiation from the companion is significant will be weakened by the decrease in atmospheric temperature gradient. L07 also suggest that there may be a similar but opposite effect in the UV spectra.  In order to determine if this is indeed the case we measured the `equivalent widths' of our CCF primary and secondary peaks at phases out of eclipse, see Figure 4.  If contributions to the UV ccfs were larger from the inner irradiated regions we would expect the primary to appear strongest at phase 0.25 and the secondary at phase 0.75.  Unlike the \ion{He}{1} lines there does not appear to be a correlation between line strength and orbital phase, which suggests that the UV lines form more or less uniformly around th stars.

It is difficult to determine the actual optical depth of formation for any particular line, but they should scale with excitation and ionization potentials.  The \ion{He}{1} lines examined by L07 are from electron transistions from the first excited level (with an excitation potential of 21 eV) and these have the largest $K$ values of $252$ and $375$ km s$^{-1}$ .  The ionization potential between \ion{He}{1} and \ion{He}{2} is slightly higher at 24.6 eV and from the L07 analysis, the $K$ values for \ion{He}{2} lines are lower at $246$ and $353$ km s$^{-1}$.  Finally, the UV spectra from which we obtain our semiamplitudes, $214.5$ and $345.8$  km s$^{-1}$, are predominantly populated with lines from \ion{Fe}{4} and \ion{Fe}{5} with ionization potentials of 30.7 and 54.8 eV.  It appears that the use of low excitation lines in contact systems where irradiation effects are present leads to overestimates of the semi-amplitude values and hence to the overall size and masses of the system.  It is clear from this analysis that the UV derived orbital elements are more representative of the true dynamic motion of the system.

\placefigure{f4}

\section{Placement in the H-R diagram and Discussion} 

Our goal in this research has been to compare the masses obtained from observations with those found from theoretical evolutionary tracks, however since both stars of TU Mus are Roche filling, we should not necessarily expect agreement with single star tracks.  In this case we will also compare the component stars to models produced from interacting binary models.  The observational masses can be determined from the combination of results from the spectroscopic and photometric orbits ($m$ sin$^{3}i$ and $i$).  Masses from evolutionary theories require the individual temperatures and luminosities of the two stars, i.e., placing the individual component stars in the H-R diagram.  We can estimate temperatures from calibrations of spectral type, and we can calculate luminosities from the observed flux ratio and the absolute magnitude of the binary (from the stellar radii associated with fits of the light curve). Individual visual magnitudes of the component stars can be obtained 
from the visual flux ratio of the two stars and the adopted $M_V$ above. Once individual visual magnitudes are obtained, we convert them 
to bolometric magnitudes and luminosities with bolometric corrections 
from \citet{How89}. Our calculated visual flux ratio is $r_{V} = 0.49 \pm 0.05$.   The luminosities and radii derived  are 
$\log L_p/L_\odot = 5.02 \pm 0.08$, $\log L_s/L_\odot = 4.55 \pm 0.09$, $R_p/R_\odot = 7.2 \pm 0.5$, and $R_s/R_\odot = 5.7 \pm 0.5$.  Our results are plotted in the H-R diagram in Figure~5 together with evolutionary tracks for single massive stars from \citet{Sch92}.  At this point in our analysis we typically compare the stellar positions with the evolutionary tracks and determine what masses they would predict.  
\placefigure{f5}

Combining the light curve model with the results from the spectroscopic orbit results in observational masses of $M_p/M_\odot = 16.8 \pm 0.4$ and $M_s/M_\odot = 10.5 \pm 0.3$.  Masses predicted from single star evolutionary tracks are significantly higher, $27.7 \pm 1.5 M_\odot$ and $18.0 \pm 1.5 M_\odot$.  Of course since both stars are in Roche lobe contact, we would not expect them to have evolved as single objects.  The severe overluminosity of these Roche filling stars is in agreement with other interacting massive binaries, specifically HD~115071 , HD~36486 , HD~35652, HD~209481, BD+66$^\circ$1521, HD~228854, HD~190967, HD~106871, and HD~35921.  These systems are listed in Table 4, along with their individual periods, mass ratios, component spectral classes, evolutionary masses from single star tracks, and dynamical masses. 

\placetable{t4}

An interacting massive binary is a complicated beast.  Trying to determine how it got to its current state is a little like trying to unspill milk.  One method is to compare its current state with other similar systems.  Interacting massive binaries fall into two broad categories: semi-detached and contact.  Its interesting to note that in all but one of the semi-detached systems in Table 4, it is the secondary that is in RLOF.  In the Case A (donor star still in core  hydrogen burning, CHB) scenario, the radius of the initially more massive star slowly reaches its Roche surface as it evolves along the main sequence.  The initially lower mass star, due to its slower evolution, is not the first to overflow.  Applying this to the semi-detached systems we infer that the current mass loser (or secondary) was initially the more massive object.  The period of a binary will shrink during mass transfer as long as the more massive star is losing mass. After enough mass has been transferred to `flip' the mass ratio, the separation between the stars will increase allowing RLOF to cease.  That we mainly see systems with the less massive object in RLOF agrees with theoretical predictions \citep[WLB01,][]{Wel01b,Pol94} that the initial transfer of mass is exceedingly quick (fast Case A), but that the turning off of slow mass transfer (slow Case A) may take a considerable amount of time. The mass ratios of these semi-detached systems range from $0.353 - 0.556$, indicating that there has been significant mass lost from the intially more massive star.  Unfortunately only one of the semi-detached systems, HD~209481, was observed by \citet{Tha97} in her survey of H$\alpha$ emission in close, O-type star binaries.  It did not have strong the H$\alpha$ emission that is associated with large RLOF gas streams.  Again this suggests that although the secondary is filling its Roche volume, there is not a significant amount of mass being transferred at this time.     

If the semi-detached systems are the expected result when the intially more massive star in a binary evolves to reach its Roche radius, what scenario leads to a contact system?  The are some obvious differences between the two groups.  The observed mass ratios for the three contact systems in Table 4 are slightly higher ($q_{av} = 0.64$) than those of the semi-detached systems ($q_{av} = 0.47$).  The components are also more closely matched, within one spectral type of each other and with similar luminosity classifications. 
WLB01 present evolutionary calculations for 74 close binary systems, with masses ranging from $26 ... 6 M_{\odot}$ for Case A and B scenarios.  On the basis of our derived luminosity classes (both stars are V) and the $\log g$ values determined from our light curve analysis ($3.98$ and $3.97$), we expect that HD~100213 is undergoing Case A evolution.  According to the WLB01 models, Case A contact systems can occur when the initial period of the system is very small and/or when the mass ratio, $q = M_2/M_1$, is either very small or very large.  For a short period system, the initially more massive star reaches its Roche radius fairly early in its evolution.  If its companion has much less mass than itself ($q << 1$) the ensuing mass transfer will proceed very rapidly as the exhange of any mass reduces the period causing the loser star to overflow at a faster rate.  The combination of the large increase in mass of the gainer star and the smaller orbit causes it to expand and reach its Roche volume.  If the companion has a mass similar to the loser star ($q \approx 1$), it will gain a much smaller amount of material from mass transfer.  In this case only a small mass transfer will `flip' the mass ratio.  However the increase in mass will cause the gainer star to `rejuvenate'.  Rejuvenation is when the star adapts its convective core mass to its new total mass \citep{Hel83,Hel84,Bra95}.  The new larger mass gainer star begins to evolve at a faster pace than its companion and reaches its Roche limit.  The state of the gainer star (original secondary) when it reaches its Roche volume determines whether the contact state is called Case AA (secondary still in CHB) or Case AB (secondary between CHB and CHeB).  In Case AB the mass transfer rate is very large as the star is rapidly increasing in radii.  This exchange  happens quickly and will increase the orbital period and separation significantly.

For HD~100213 we suggest the following scenario. The two stars began with a mass ratio very close to unity and a period slightly smaller than the current value.  The actual initial masses and period depend upon how much of the lost mass is transferred.  The initially more massive star reaches its Roche volume and mass is transferred to the gainer star.  The stars move closer together for a brief time until the gainer star is more massive.  Mass transfered after this point causes the stars to move further from one another, however this does not cease the RLOF. Fast Case A mass transfer continues and doesn't end until the loser star is much less massive.  WLB01 explain that since the $M_{core}/M_{envelope}$ is increasing for the loser star this causes its radius to increase.  This agrees well with our observed mass ratio of $0.625$.  Finally the gainer star expands due to the increase in mass and its rejuvenated interior until it also reaches its Roche volume.  This is current status of the binary.  What is the current state of mass transfer?  \citet{May04} examined the historical records \citep[AG75, ST95, T03]{Oos28,Oos30,Kni71} and argue that the period of TU Mus lengthens, but not regularly.  Another study by \citet{Qia07} found a long term increase in period along with superimposed cyclical variations.  The cyclical variations they attribute to the presence of a nearby ($\approx 50$ AU), low mass ($\approx 2 M_\odot$) tertiary companion.  They argue that the long term increase in the observed periods is consistent with a conservative mass transfer rate of $4.2\times 10^{-7} M_\odot$ yr$^{-1}$ from the lower mass secondary to the primary. \citet{Tha97} observed HD~100213 looking for H$\alpha$ emission, which we associate with mass transfer, and detected none.  We suggest that the system is just barely in contact, with both stars still in a CHB stage, Case AA.

We highly recommend that this rare contact system, now with well determined masses, radii, separation, and temperatures be modelled using binary evolutionary codes.  \citet{Pol94} also examined Case A and B interacting systems and determined that those with small initial periods and small or large mass ratios will reach contact.  However due to his differing treatment of convection, \citet{Pol94} predicts that all gainer stars will reach contact through Case AB.  All gainer stars are quickly rejuvenated to the point that they reach the end of CHB and then expand rapidly to contact.  Another issue this system can address is how much mass lost from the loser is accreted upon the gainer.  Both \citet{Pol94} and WLB01 have treated the mass transfer as conservative, i.e., all mass lost by the donor is accreted by the gainer. The mechanism for non-conservative mass transfer is the rapid spinning up of the gainer star until it reaches its critical rotation velocity \citep{Wel01a,Langer03,Langer04,Pet04}.  Neither star in this system is near critical rotation, which suggests that the mass transfer has been conservative. What \citet{Pol94} and WLB01 agree upon is the prediction that Case A contact systems will end in a binary merger.  Both find that the resulting periods are too small for successful common envelope ejection.  A merger results in a very exotic object, which may be a precursor to Gamma Ray Burster or other strange supernova. 

\acknowledgments
We are very grateful for the electronic version of AG75 photometric data provided by Jens Viggo Clausen and Johannes Andersen.  The {\it IUE} data presented in this paper were obtained from the Multimission Archive at the Space Telescope Science Institute (MAST).  STScI is operated by the Association of Universities for Research in Astronomy, Inc., under NASA contract NAS 5-26555.  Support for MAST for non-HST data is provided by the NASA Office of Space Science via grant NAG 5-7584 and by other grants and contracts.  This research has made use of the SIMBAD database,    
operated at CDS, Strasbourg, France.  This research has also made use of the ESA, HIPPARCOS and TYCHO catalogues.
Support for this work was provided by NASA grant NAG 5-2979.
Institutional support for L.R.P. has been provided from the College of Charleston School of Sciences and Mathematics.  Additional support for L.R.P.\ was provided from AR Nos. 09945 and 11275 from STScI, program GO63 from the Far Ultraviolet Spectrographic Explorer ({\it FUSE}), NASA 06-ADP06-68, and NSF grant AST-0506541.  We gratefully acknowledge all this support.



\newpage

\centerline{Figure Captions}
\vskip 0.1 in

\figcaption[fig1.ps]{Radial velocity measurements for the primary ({\it filled circles}) and 
secondary ({\it open circles}) based on {\it IUE} spectra.   Solid lines are drawn
 for both the primary and secondary radial velocity solutions. 
 \label{f1}}
\vskip 0.1 in

\figcaption[fig2.ps]{The observed {\it HIPPARCOS} light curve ({\it filled circles}) and AG75 differential $V$ light curve ({\it filled triangles} offset by +5.74 mag) plotted together with the optimum model fit (after adjusting the stellar effective temperatures, {\it solid line}). \label{f2}}

\figcaption[fig3.ps]{The observed {\it HIPPARCOS} and AG75 differential $V$ light curves (offset by +5.74) plotted with our minimum acceptable inclination model ({\it short dashes}), maximum acceptable inclination model ({\it long dashes}) and optimal model ({\it solid line}).  Only the secondary eclipse is shown. The data points are greatly reduced in size to better see the model fits. \label{f3}}  
\vskip 0.1 in

\figcaption[fig4.ps]{Measured `equivalent widths' of our CCFs outside of eclipses for the primary ({\it filled circles}) and secondary ({\it open circles}). \label{f6}}

\figcaption[fig5.ps]{A HR diagram of the binary system TU Mus.  The filled (open) circle represents the primary (secondary) component.  The overdrawn evolutionary tracks are for single stars from Schaller et al. (1992). \label{f7}}

\newpage

\begin{deluxetable}{rrcrrrrrr}
\footnotesize
\tablewidth{0pc}
\tablenum{1}
\tablecaption{IUE Radial Velocity Measurements \label{t1}}
\tablehead{
  \colhead{SWP}     &\colhead{HJD}         
 &\colhead{Orbital}     &\colhead{$V_1$}   &\colhead{$(O--C)_1$}       
 &\colhead{}
&\colhead{$V_2$}   &\colhead{$(O--C)_2$}
&\colhead{} \\
\vspace{-10pt} \\
  \colhead{Image}   &\colhead{(-2,400,000)} 
 &\colhead{Phase} &\colhead{(km s$^{-1}$)} &\colhead{(km s$^{-1}$)}
 &\colhead{$W_1$} 
&\colhead{(km s$^{-1}$)}&\colhead{(km s$^{-1}$)} &\colhead{$W_2$}}
\startdata
15916 & 44970.915 & 0.134 & 154.4 &   20.5 & 0 & --214.3 &  11.6 & 0\\
19642 & 45429.986 & 0.048 & 199.5 &   3.9  & 1 & --330.0 &  --4.6 & 1\\
19643 & 45430.049 & 0.093 &  183.4 &  13.9  & 1 & --274.3 &  9.0 & 1\\
19651 & 45430.480 & 0.404 & --207.5 & --21.7  & 1 &  284.8 &  --4.7 & 1\\
37879 & 47883.732 & 0.792 &  92.0 &  45.6  & 0 &  --119.2  &  --34.3 & 0\\
37880 & 47883.809 & 0.847 & 147.8 &  33.9  & 0 & --241.1 &  --47.5 & 0\\

37906 & 47887.590 & 0.573 & --194.3 &  7.5  & 1 &  313.0 &  --2.2 & 1\\
37907 & 47887.654 & 0.619 & --171.9 &  --5.3  & 1 & 282.4 &  23.8 & 1\\
54353 & 49817.760 & 0.906 & 165.3&  --3.2  & 1 & --302.3 &  --20.7 & 1\\
54356 & 49817.933 & 0.030 & 191.3 &  --10.2  & 1 & --325.7  &   9.0 & 1\\
\vspace{-7pt} \\
54360 & 49818.117 & 0.163 & 130.6 &  28.1  & 0 & --234.6  & --59.3 & 0\\
54371 & 49818.514 & 0.449 &  --206.8 & 6.0  & 1 & 311.9  &  --21.2 & 1\\
54374 & 49818.700 & 0.583 &  --189.1 & 6.2  & 1 & 316.2  &   11.4 & 1\\
54378 & 49818.917 & 0.739 &  --64.1 &  --40.5  & 0 & 64.6  &  36.4 & 0\\
54382 & 49819.097 & 0.869 &  152.7 &   16.0  & 0 & --284.6  & --54.2 & 0\\
54390 & 49819.481 & 0.146 &  143.8 &  22.6  & 0 & --244.4 &  --39.0 & 0\\
54394 & 49819.667 & 0.280 &  --126.3 & --76.7  & 0 & 83.1 & 13.3 & 0\\
54397 & 49819.813 & 0.385 & --194.7 &   --24.3  & 0 & 274.7  &  10.0 & 0\\
54407 & 49820.633 & 0.976 & 200.0 & --2.9  & 1 & --332.3  &  4.7 & 1\\
54411 & 49820.805 & 0.100 & 167.5 &   3.5  & 0 & --276.5  &   --2.1 & 0\\
\vspace{-7pt} \\
54415 & 49820.994 & 0.237 & 42.1 &  33.3  & 0 & --87.4  & --63.2 & 0\\
54419 & 49821.185 & 0.374 & --181.9 &  --21.6  & 0 & 271.0  & 22.7 & 0\\
54426 & 49821.471 & 0.580 & --191.7 &  5.3  & 1 & 303.9  & --3.6 & 1\\
\enddata
\end{deluxetable}

\newpage

\begin{deluxetable}{lrrrr}
\small
\tablewidth{0pc}
\tablenum{2}
\tablecaption{Circular Orbital Elements  \label{t2}}
\tablehead{
\colhead{Element} & \colhead{ST95} & \colhead{T03} &
\colhead{Primary only} & \colhead{Secondary only}}
\startdata
$P$~(days)                   & 1.3872827(17)& 1.38728653(2) & 1.387282(7) & 1.387280(6)\\
\noalign{\medskip}
$T_0$ (HJD 2,400,000+)         & 49817.879(3) & 49817.8820(1) & 49817.662(5) & 49817.666(5) \\
\noalign{\medskip}
$K_1$ (km s$^{-1}$)          & 216.7(2.7)    & 249.0(2.8)  & 214.5(2.4)  & \nodata      \\
$K_2$ (km s$^{-1}$)          & 345.4(3.1)    & 382.6(4.3)  & \nodata     & 345.8(3.2)   \\
\noalign{\medskip}
$V_{o}$ (km s$^{-1}$)        & --12.7(2.1)    & --4(4)    & \nodata     & \nodata      \\
$V_{o~1}$ (km s$^{-1}$)      & --11.8(2.1)    & \nodata  & --9.3(2.1)   & \nodata      \\
$V_{o~2}$ (km s$^{-1}$)      & --14.2(3.6)    & \nodata    & \nodata     & 4.9(2.9)     \\
\noalign{\medskip}
$m_1$ sin$^{3}i$ ($M_\odot$) & 15.70(34)     & 21.9(8)     & 15.64(44)     & \nodata      \\
$m_2$ sin$^{3}i$ ($M_\odot$) & 9.85(24)      & 14.3(4)     & \nodata     & 9.71(32)      \\
\noalign{\medskip}
$a_1$ sin $i$ ($R_\odot$) & 5.94(7)     & 6.97(8)   & 5.88(7)   & \nodata      \\
$a_2$ sin $i$ ($R_\odot$) & 9.47(8)     & 10.5(1)   & \nodata     & 9.48(9)    \\
\noalign{\medskip}
r.m.s. (km s$^{-1}$)         & 13.0           & \nodata     & \nodata     & \nodata      \\
r.m.s.$_1$ (km s$^{-1}$)     & \nodata       & \nodata      & 10.1        & \nodata      \\
r.m.s.$_2$ (km s$^{-1}$)     & \nodata       & \nodata     & \nodata     & 13.7         \\
\enddata
\end{deluxetable}

\newpage
\begin{deluxetable}{ccccccc}
\small
\tablewidth{0pc}
\tablenum{3}
\tablecaption{$V_r$ from Synthetic Profiles  \label{t3}}
\tablehead{
\colhead{} & \colhead{} & \colhead{} & \colhead{UV $V_{r,p}$} &
\colhead{UV $V_{r,s}$} & \colhead{Optical $V_{r,p}$} &
\colhead{Optical $V_{r,s}$} \\
\vspace{-11pt} \\
\colhead{Case} & \colhead{Reflection Effects?} & \colhead{Line Variability?} & \colhead{(km s$^{-1}$)} & \colhead{(km s$^{-1}$)} & \colhead{(km s$^{-1}$)} & \colhead{(km s$^{-1}$)}}
\startdata
Geometric & \nodata & \nodata & --214.5 & 345.8 & --214.5 & 345.8\\
1 & off & off & --210.9 & 344.9 & --209.9 & 345.1\\
2 & on & off & --210.6 & 343.1 & --209.5 & 343.7\\
3 & on & on & --211.6 & 343.3 & --209.7 & 345.2\\
\enddata
\end{deluxetable}
\newpage

\begin{deluxetable}{lcccccccccc}
\rotate
\tablewidth{0pt}
\tabletypesize{\scriptsize}
\tablenum{4}
\tablecaption{Interacting OB Star Binaries \label{t4}}
\tablehead{
 \colhead{Star}               &
 \colhead{$P$} &
 \colhead{$q$}  &
 \colhead{Primary}  &
 \colhead{Secondary} &
 \multispan{2}\hfil M$_{evol}$ (M$_{\odot}$) \hfil  &
 \multispan{2}\hfil M$_{dyn}$ (M$_{\odot}$) \hfil  &
 \colhead{Roche Filling}   &
 \colhead{} \\
\vspace{-11pt} \\
\colhead{Name}   &
\colhead{(days)}  &
\colhead{($M_2/M_1$)}   &
\colhead{Type}  &
\colhead{Type}   &
\colhead{Primary}  &
\colhead{Secondary}  &
\colhead{Primary}   &
\colhead{Secondary}  &
\colhead{Star?}  &
\colhead{Reference}  }
\startdata
\multispan{11}\hfil Semi-Detached Systems \hfil \\
\multispan{11}\hrulefill \\
HD~35652 = IU Aur & 1.81 & 0.503 & B0.5 V & B0.5 V & $~15$ & $~13$ & 14.5 & 7.3 & Secondary & 2 \\
HD~115071 = V961 Cen & 2.73 & 0.575 & O9.5 V & B0.2 III & $~18$ & $~15$ & 11.6 & 6.7 & Secondary & 4 \\
HD~209481 = LZ Cep & 3.07 & 0.415 & O8.5 III? & O9.5-B0 V? & $~21$ & $~18$ & 15 & 6.3 & Secondary & 2 \\
HD~106871 = AB Cru & 3.41 & 0.353 & O8 V & B0.5 ? & $~31$ & $~16$ & 19.8 & 7.0 & Secondary & 3 \\
BD~+66$^\circ$1521 = XZ Cep & 5.10 & 0.405 & O9.5 V & B1 III & $~15$ & $~14$ & 15.8 & 6.4 & Secondary & 1 \\
HD~36486 = $\delta$ Ori A & 5.73 & 0.500 & O9.5 II & B0.5 III & $~28$ & $~12$ & 11.2 & 5.6 & Primary? & 5 \\
HD~190967 = V448 Cyg & 6.52 & 0.556 & O9.5 V & B1 II-Ia & $~15$ & $~14$ & 25.2 & 14 & Secondary & 1 \\
\vspace{-1pt} \\
\multispan{11}\hfil Contact Systems \hfil \\
\multispan{11}\hrulefill \\
HD~100213 = TU Mus & 1.38 & 0.652 & O7 V & O8 V & $~28$ & $~18$ & 16.8 & 10.5 & Both & 10 \\
HD~228854 = V382 Cyg & 1.89 & 0.742 & O7.3 V & O7.7 V & $~32$ & $~24$ & 26 & 19.3 & Both & 1 \\
HD~35921 = LY Aur & 4.00 & 0.528 & O9 III & O9.5 III & $~34$ & $~28$ & 24 & 12.7 & Both & 6,7,8,9 \\
\enddata
\tablecaption{REFERENCES.-(1) Harries, Hilditch, \& Hill (1997); (2) Harries, Hilditch, \& Hill (1998); (3) Lorenz, Mayer, \& Drechsel (1994); (4) Penny et al. (2002); (5) Harvin et al. (2002); (6) Stickland et al. (1994); (7) Drechsel, Lorenz, \& Mayer (1989); (8) Popper (1982); (9) Andersen, Batten, \& Hilditch (1974); (10) this paper.}
\end{deluxetable}

\newpage

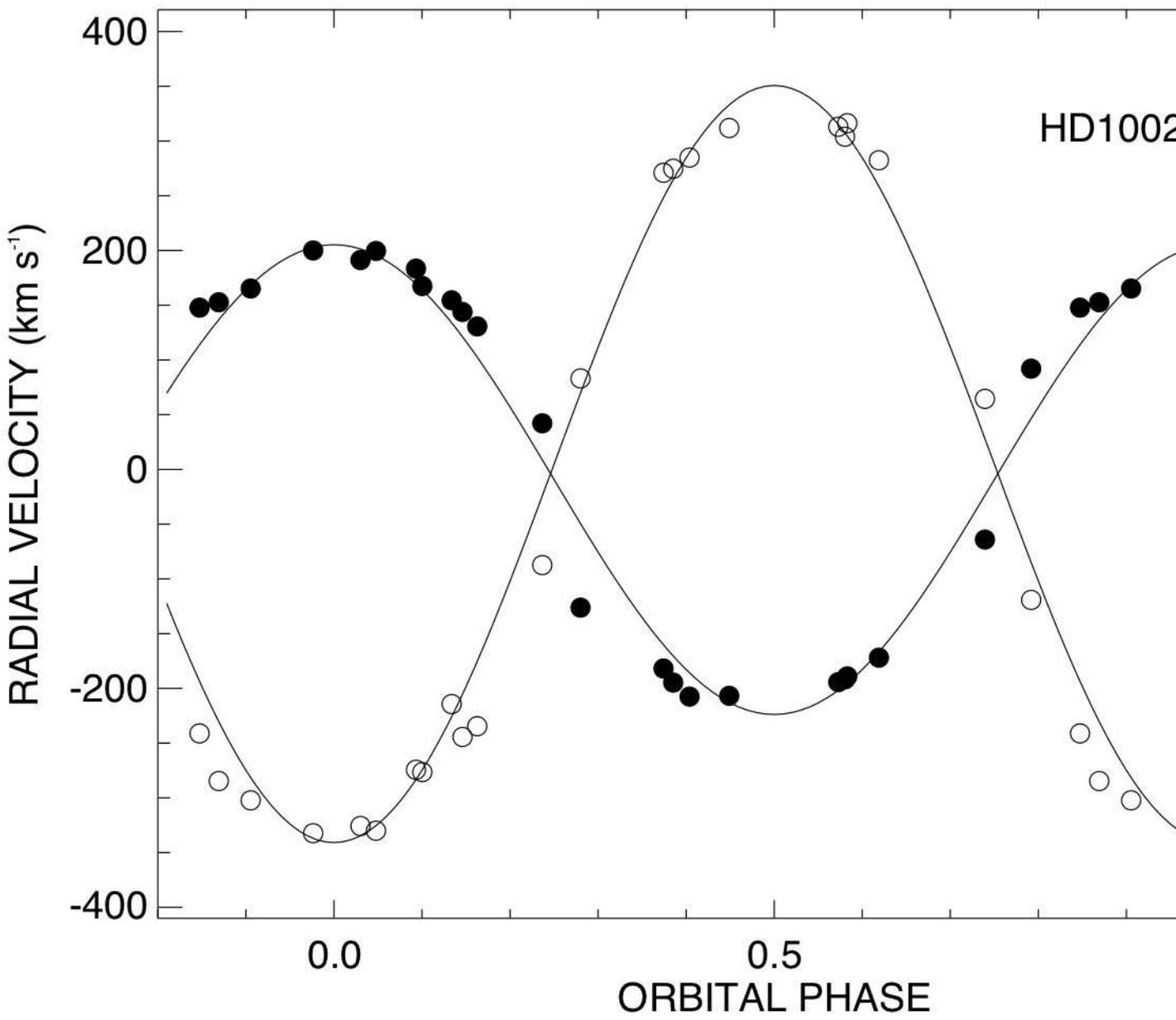
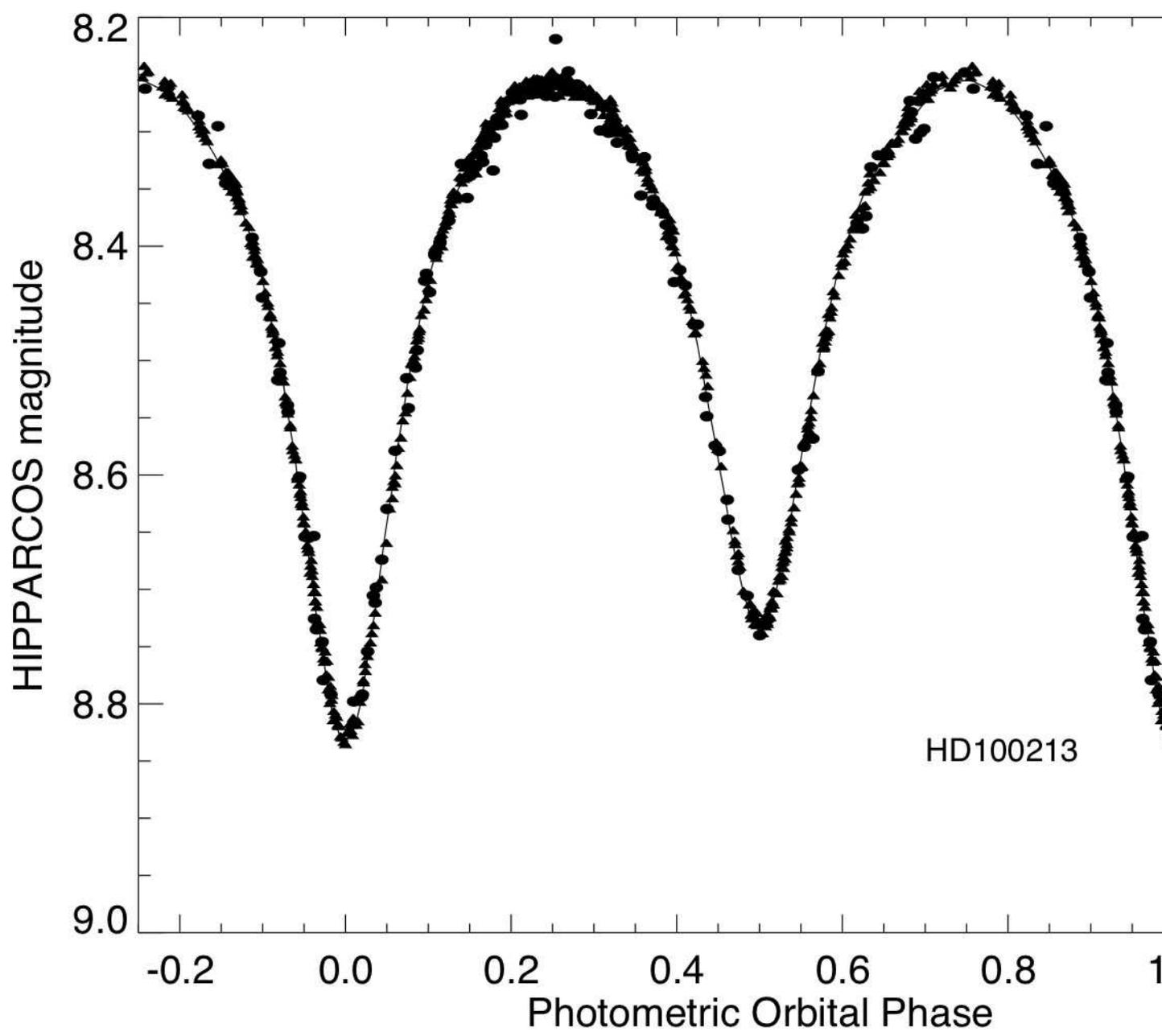
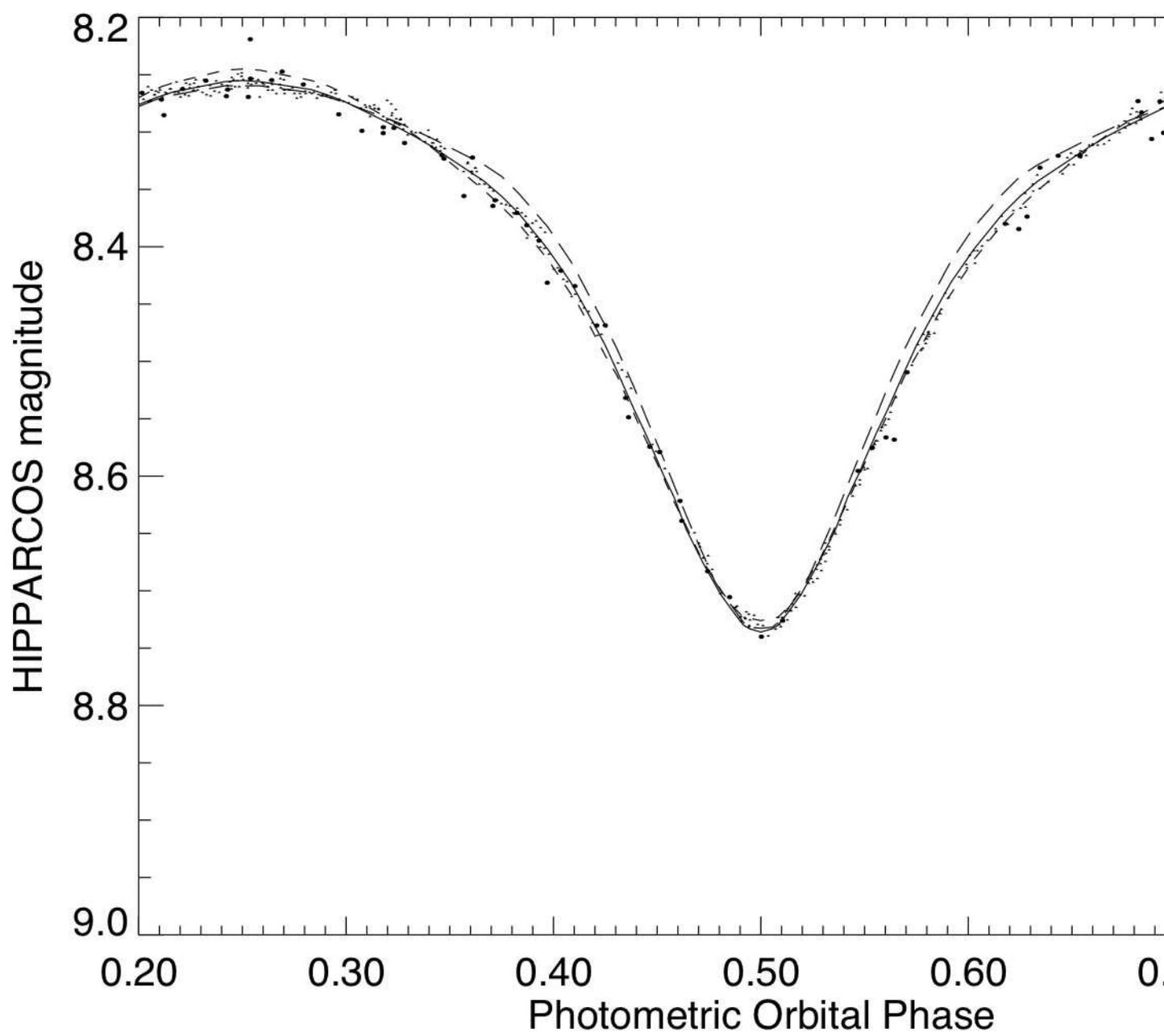
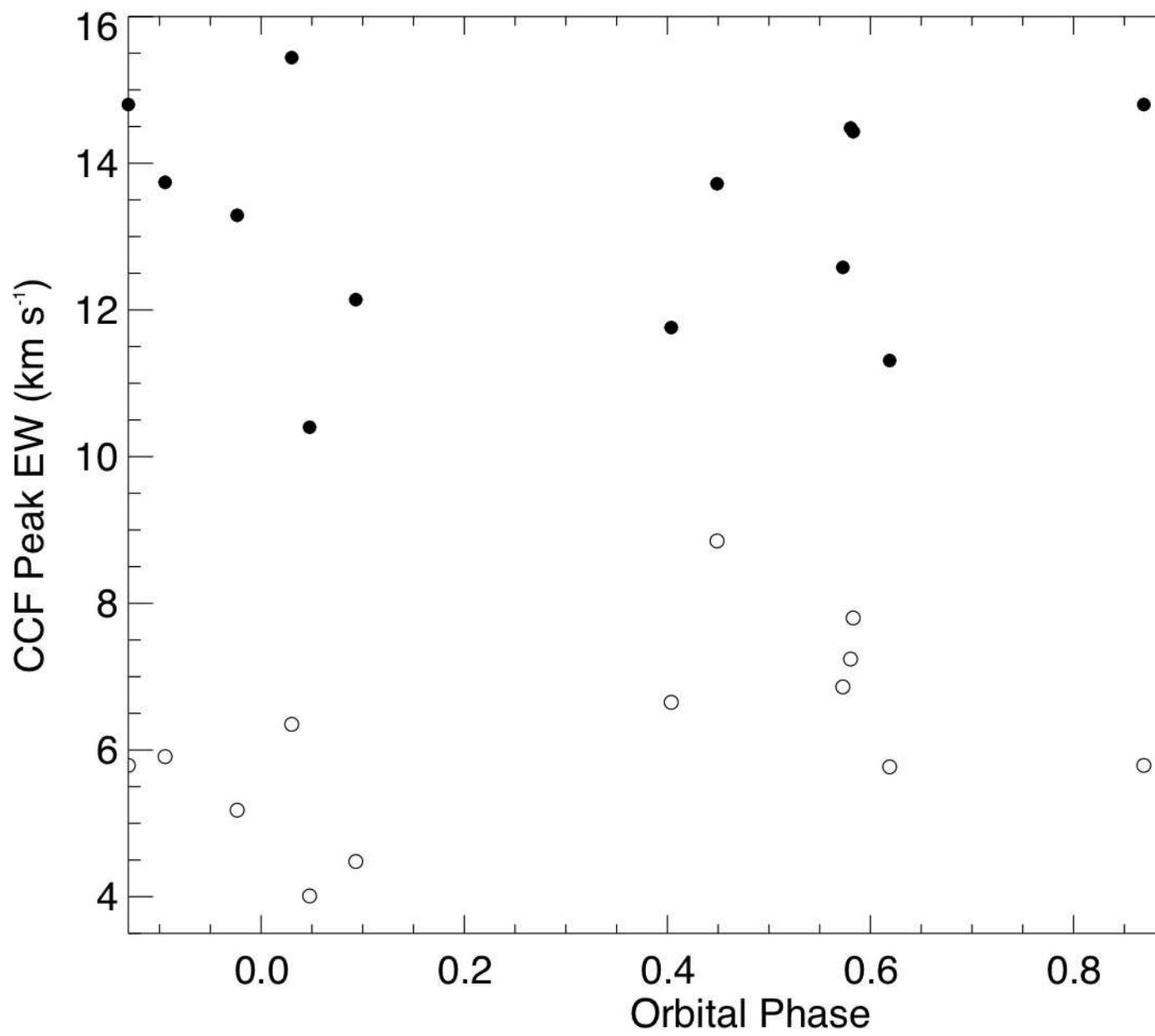
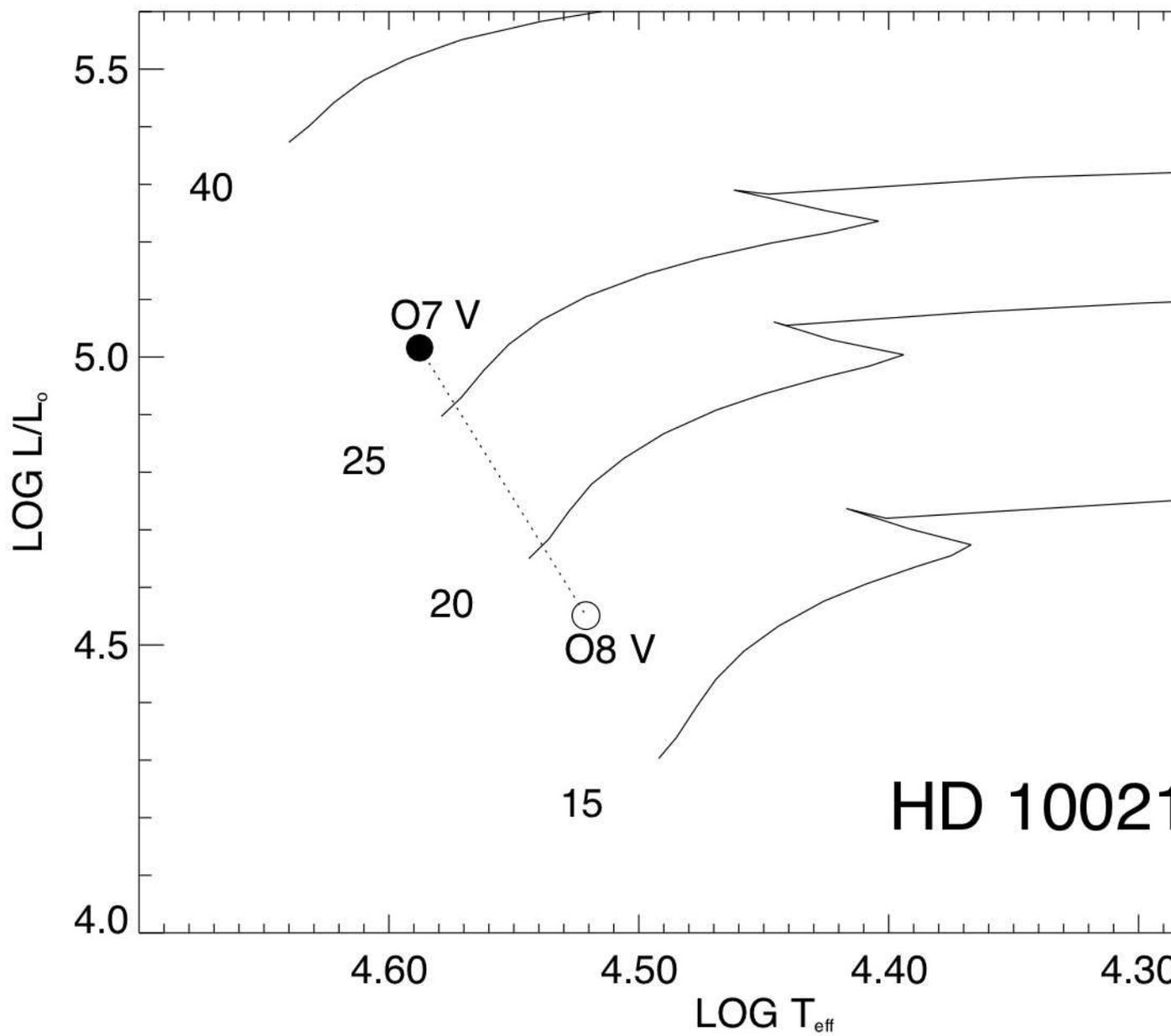
\end{document}